\documentclass[14pt]{article}

\begin{document}

\begin{center}
CRITICAL VALUES OF THE EXTERNAL MAGNETIC FIELD LEADING BIOLOGICAL EFFECTS IN THE
HUMAN ORGANISM
\end{center}

\begin{center}
Z. Kanokov and Sh. Z. Kanokova
\end{center}

\begin{center}
National University of Uzbekistan, Tashkent, Uzbekistan
\end{center}

\begin{abstract}
In the framework of the simplified stochastic model the
critical values of an induction of the external magnetic field
leading to sharp increase of fluctuations of a casual current of
biologically important ions in different blood vessels of a human
body are calculated.
\end{abstract}

Interest to the magnetic-biology is caused, first of all, by the
ecological reasons. Present time exist sets of the experimental
facts, that natural and technogenic weak, low-frequency electric and
magnetic fields (LFEMF) represent potential threat for health of
people and they are not less essential ecological factor, than other
meteorological ones. With development of a radio communication, TV,
radionavigation, a radar-location and mobile telephones in front of
mankind it arises a problem of the environment ``electromagnetic
pollution'' which results from activity of people. Thus, biosphere
of our planet feels on itself the regular influence of fields of the
artificial origin besides the natural geomagnetic and space
electromagnetic field. At interaction a weak electromagnetic
radiation with biological objects the integrated heating does not
exceed 0.1 degrees [1,2]. Therefore, earlier was considered, that
LFEMF are safe for the person, biological action of such fields
seemed to be impossible from the point of view of physics. The
experimental researches, which were performed for the last decades,
testify the high sensitivity of biological systems to the LFEMF of
weak intensity [3-5]. However, the physical, biological and chemical
mechanisms of influence of the given fields on biological systems in
all spent researches were not established well. It should be
stressed that the results of `different experimental groups often
have inconsistent character, that essentially complicates
understanding of mechanisms of action of the weak LFEMF on the
biological objects [5-7]. Thus, most actual questions in
electromagnetobiology are:

\begin{itemize}
    \item revealing of the basic laws of interaction of an electromagnetic field of a
technogenic origin with biological systems and an environment;
    \item development of reliable means of protection against negative influence of
LFEMFon ecological systems.
\end{itemize}

It is well-known, that biological objects are the complex open
nonlinear dynamic systems and their states can be determining in
their reaction and not just the influencing external fields. Their
complexity is caused by that, being macroscopically, they consist of
many objects accepted for elements of their structure. By
consideration of mechanisms of interaction of the weak LFEMF with
elements of complex biological system there is a necessity of search
of the most fundamental principles determining such interaction.
Apparently, these principles lay in the field of studying
non-equilibrium and non-linear interactions of biological systems as
a whole, or their separate elements. Prigozhin   marked that open
systems continuously fluctuate [8]. Fluctuation is casual deviation
of the physical quantities describing system from their mean values.
Sometimes a separate fluctuation or their combination can become so
strong, that the organization existed before does not survive and
collapses. Therefore, the huge interest represents not only research
of interaction of an electromagnetic field with biological systems
as a whole, but also its interaction with separate elements of
system. It is possible to carry vessels, cells, lipids, fibers, and
also their water solutions to the last.

In Refs [9,10] the simplified stochastic model to study influence of
a weak external magnetic and electric field on fluctuation of an
ionic casual current in blood has been offered. It is shown, that
the external stationary magnetic field can cause sharp increase in
fluctuation of an ionic electric casual current in biological cells,
in particular in blood vessels owing to Brown movement of the free
charged ions.Analytical expression for kinetic energy of a molecule
of environment in the considered volume was obtained. The results of
energy of molecules in the volume of a capillary and an aorta near
to critical value of an external magnetic field were presented. It
is shown, that as approaching critical values of a magnetic
field,the averaged energy of the molecule localized in a capillary
increases for some orders of magnitude in comparison with its
thermal energy. As a result the pressure of blood can increase in
the considered volume which may appears as a small red macula on the
skin of man body.  As well as the large value of energy may be
enough for destruction of chemical communications. Even if the value
of a magnetic field not so close to critical values, a significant
effect can be reached with increase in time of an exposition of a
magnetic field. In Refs. [9,10] all numerical calculations have been
performed only for ions of calcium-Ca$^{2+}$. In the alive
organisms, except for ions Ca$^{2+}$ also ions of magnesium
Mg$^{2 +}$, potassium K$^{+}$, sodium Na$^{ +}$, iron Fe$^{2+}$ are
biologically important.

In this work we have calculated critical values $B_{cr }$ of an
induction of external magnetic field leading to sharp increase
fluctuations of a casual current of biologically important ions in
different blood vessels of a human body (Table 1). Thus we consider
each blood vessel as a separate element of an organism and $B_{\rm cr
}$ is calculated by the formula which easily it turns out from
expression resulted in Ref.[9,10]

\begin{center}

\[
B_{cr}=\frac{6\pi\eta r}{qn}
\]

\end{center}

{\raggedright where
$\eta=(1.1\div{} 1,2) \cdot 10^{-3}$ kg/(m$\cdot$s) is the
viscosity coefficient of blood, $r$ is the radius of an ion,
$q$ is the charge of an ion,  $n$ is a number of ions of the given
element volume  $\ V=\frac{\pi{}d^2}{4}\cdot{}l\ $ of a vessel.}

\begin{center}
{\small Table 1. Geometrical sizes of the blood vessel of human body.
\vspace{3pt} \noindent
\begin{tabular}{|c|c|c|c|c|c|}
\hline
 Vessel &  Diameter, $d$   & Cross section, $S$ &  Length, $l$ &
Total number  &  Volume of
vessel, $V$
 \\
 & ($10^{-2}$m) & ($10^{-4}$ m$^{2}$) & $(10^{-2}$ m) &
in the organism &   ($10^{-8}$ m$^3$) \\
\hline
Aorta
 &  3.2-1.6 & 8.0-2.00 &  80  &  1 &  640-160
 \\
\hline
 Hollow veins & 2.0 & 3.14 & 50 & 2 & 157 \\
\hline
 Large veins &  1.0-0.5 &  0.80-0.20 &  30-10 &  1000 & 24-2 \\
\hline  Large artery &  0.6-0.1 &  0.28-0.01
 &  40-20 &  1000 &  11.2-0.2
 \\
\hline
 Small artery  & 0.1-0.02& 0.01-0.003 &  5-0.2 &
 10$^{8}$ &  5 $\cdot 10^{-2}6\cdot10^{-4}$ \\
\hline
 Capillary &  (10-5)$\cdot 10^{-4}$ & $(19.63\div78.5)\cdot 10^{-8}$
 &  0.1 &  10$^{9}$ &  $(2.0\div7.9)\cdot 10^{-8}$  \\
\hline
\end{tabular}
}\vspace{2pt}

\end{center}

The number of ions in the considered volume was determined by the
use of data presented in Table 2. Results of calculations are presented in Table 3.

\begin{center}
Table 2. Parameters of a mineral exchange in blood[11].

\vspace{3pt} \noindent
\begin{tabular}{|c|c|}
\hline
 Parameter &  Mole /liter \\
\hline
 Calcium in whey of blood &  2.25-3
 \\
\hline
 Magnesium in whey of blood &
 0.70-0.99 \\
\hline
 Iron ie whey of blood &  12.5-30.4
 \\
\hline
 Potassium in plasmas of blood
 &  3.48-5.3 \\
\hline
 Sodium in plasmas of blood &
 130.5-156.6 \\
\hline
\end{tabular}
\vspace{2pt}
\end{center}

 Table 3. Critical values of induction $B_{\rm cr}$
an external magnetic field (T).

\vspace{3pt} \noindent
\begin{tabular}{|c|c|c|c|c|c|c|}
\hline
 Element of
 & Aorta  &  Hollow veins
  &  Large veins
 &  Large artery
  &  Small artery
 &  Capillary \\
vessel
 &10$^{-12}$& 10$^{-12}$ & 10$^{-12}$ & 10$^{-12}$& 10$^{-10}$ & 10$^{-3}$\\
\hline
 B(Ca$^{2+}$)
 &  0.50-0.88 &  5.0 &  0.2-1.0 &
  0.7-23.5 &  2.3-70.6 &  0.26 \\
\hline
 B(K$^{+}$)  &  0.15-0.26
 &  0.26 &  1.6-21
 &  7-184
 &  45-9500
 &  0.6 \\
\hline
 B(Na$^{+}$)
 &  0.32.-0.54
 &  0.6 &  4.3-65 &  21-650 &  650-16250 &  1.4 \\
\hline
 B(Mg$^{2+}$)
 &  0.12-0.32 &  0.4 &  2.6-50
 &  13-435 &  435-12500 &  0.9 \\
\hline
 B(Fe$^{2+}$) &  2.3-3.5 &  3.6 &  27-280
 &  140-4030 & 450-9500 &  9.4 \\
\hline
\end{tabular}
\vspace{2pt}
\\

{\bf Conclusions}

In this work critical values of an induction of an external magnetic
field of $B$$_{cr}$, which influence on ions of the vital elements
can cause change in a biological organism, for the first time have
been calculated. Results of calculation show that the values of
$B$$_{cr}$ depending on tho volume a vessel and number of ions rn an
important biological element tn the given volume accepts values from
10$^{-12}$T us to 10$^{-3}$T. These results allow us to explain
the inconsistency and discrepancy between different experimental
data obtained by different experimental groups at various times: any
biological organism represents itself as complex and individual
system for each object. Therefore, the data corresponding to one
person cannot be considered as the standard for another person. The
everything depends on structure of blood, concentration of the vital
elements and characteristics of vessels in an organism. The consent
of ions of biological important elements in an organism is various
depending on the nature of a structure of the person. In this
connection, it will be unequivocally impossible to establish what
the biological response to all organisms as a whole.
\\

\end{document}